\documentclass[preprint,prb,preprintnumbers,amsmath,amssymb]{revtex4}

\usepackage{graphicx}
\usepackage{bm}

\begin{document}
\title{Triplet absorption in carbon nanotubes: a TD-DFT study}

\author {Sergei \surname{Tretiak}}
\email{serg@lanl.gov} \affiliation{Theoretical Division, and Center for Integrated Nanotechnologies
(CINT), Los Alamos National Laboratory, Los Alamos, NM 87545}

\begin{abstract}

We predict properties of triplet excited states in single-walled carbon nanotubes (CNTs) using a
time-dependent density-functional theory (TD-DFT).  We show that the lowest triplet state energy in
CNTs to be about 0.2-0.3 eV lower than the lowest singlet states. Like in $\pi$-conjugated
polymers, the lowest CNT triplets are spatially localized. These states show strong optical
absorption at about 0.5-0.6 eV to the higher lying delocalized triplet states. These results
demonstrate striking similarity of the electronic features between CNTs and $\pi$-conjugated
polymers and provide explicit guidelines for spectroscopic detection of CNT triplet states.

\end{abstract}

\maketitle

\begin{figure*}
   \includegraphics[width=0.9\textwidth]{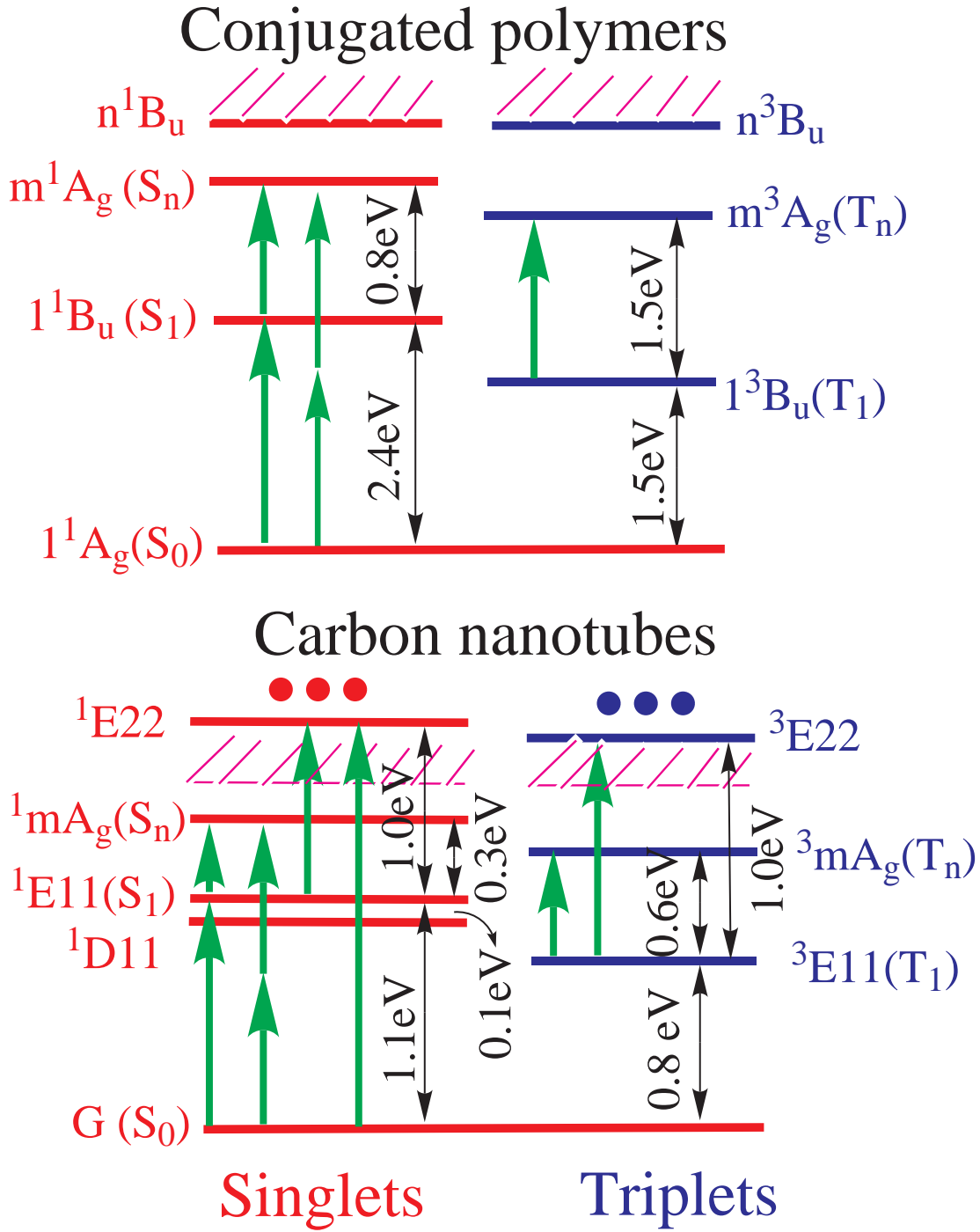}\vspace{0.3cm}
   \caption{Scheme of the excited state structure of $\pi$-conjugated polymers and
   semiconductor carbon nanotubes. Shown by grey arrows are optically allowed electronic transitions and
   typical related energy numbers corresponding to poly-phenylene-vinylene (PPV) and 1 nm diameter tube.
   Shaded area denote the beginning of a continuum for each excitonic manifold.}
   \label{fig:levels}
   \end{figure*}

Semiconductor or metal-like electronic features of single-walled carbon nanotubes (CNTs) result
from a delocalized $\pi$-electron system confined by a quasi-one-dimensional structure of the
material \cite{saito}. Understanding, probing and controlling fundamental electronic properties of
CNTs are the keys opening many exciting nanotechnology applications as nanoelectronic and
nanophotonic devices \cite{PostmaHWC:Carnst,ChenZH:Anilc}. In the past 2-3 years we have observed a
revolution in experimental studies related to carbon nanotube photophysics. Experimental
\cite{BachiloSM:Stross,O'ConnellMJ:Bangfi,KorovyankoOJ:Ultses,WangF:Theorc} and theoretical
\cite{AndoT:Exccn,ZhaoHB:Eleieo,PerebeinosVasili:Scaecn,PerebeinosV:Radlec,SpataruCatalinD.:ExcEaO,TretiakS:Excapd,araujo:067401}
work has revealed the importance of excitonic affects and electron-electron correlations enhanced
by one-dimensional confinement conditions. Moreover, recent research has demonstrated substantial
electron-phonon coupling (vibrational effects) in CNTs as well
\cite{PerebeinosV:Effecc,PlentzF:Direee,HabenichtBF:TimaiS,lanzani_tretiak,shreve:037405}. All
these phenomena are typical features of quasi-one-dimensional materials \cite{ScholesGD:Excns} such
as conjugated organic \cite{TretiakS:Denmas} and organometallic \cite{BatistaER:ExcLoc} polymers,
and mix-valence chains \cite{DexheimerSL:Femvds}. In particular, recent studies reveal many common
electronic features observed in spectra of CNTs and conjugated polymers
\cite{ZhaoHB:Eleieo,WangF:Theorc,ZhaoH.:Phoeqo,OsterbackaR:Excpal,TretiakS:Excapd}.
Figure~\ref{fig:levels} schematically shows the essential electronic states contributing to
photophysics of these materials providing common comparison baseline. We use shorthand labeling of
the relevant ground ($S_0$) and excited states ($S_1$, $T_1$, $\ldots$), which is common for
molecular physics, and field-specific notations.

So far, extensive amount of work in nanotube science has been done on the excitonic properties of
the singlet manifold only
\cite{BachiloSM:Stross,O'ConnellMJ:Bangfi,KorovyankoOJ:Ultses,WangF:Theorc,AndoT:Exccn,ZhaoHB:Eleieo,PerebeinosVasili:Scaecn,PerebeinosV:Radlec,SpataruCatalinD.:ExcEaO}.
Triplet states in CNTs have not been detected experimentally yet. Properties of triplet states in
CNTs should affect a number of fundamental physics phenomena. Singlet-triplet splitting in
low-dimensional materials is a measure of electronic correlation strength and exchange effects,
similar to that of exciton binding energy shows. It is not known how relaxation of photoexcitations
occurs in CNTs \cite{KorovyankoOJ:Ultses,lanzani_tretiak}. Carbon nanotubes are weakly-emissive
materials. Recent studies attribute this property to the 'dark' singlet excitons below the
optically allowed states \cite{ZhaoHB:Eleieo,ZhaoH.:Phoeqo,CapazRB:Diaacd}. It is remains to be
seen if the triplet states play any substantial role in the non-radiative decay of
photoexcitations. Note that in spite of weak spin-orbit coupling, a substantial fraction of
photoexcitations undergoes intersystem crossing to triplets in such related systems as fullerenes.
Likewise, photoexcitations to the higher lying CNTs excitons (e.g. E22, E33) generate unbound
electron-hole pairs, which can recombine back to excitons formally at a statistical ratio (three
triplet states are formed per one singlet). Subsequently, CNTs triplet states can play even more
important role for relaxation processes from the higher excited states
\cite{ZhaoH.:Phoeqo,lanzani_tretiak} or for the dynamics of electron and hole carriers in transport
processes \cite{PostmaHWC:Carnst,ChenZH:Anilc}. Moreover, triplet states may be involved in a
number of other important fundamental phenomena such as impact ionization and electro-luminescence.

In this letter we report extensive computational modeling of excited states of CNTs using first
principles based time-dependent density functional theory (TD-DFT). First of all, we ensure that
the hybrid DFT used is quantitatively accurate for both singlet and triplet states in conjugated
polymers and singlet states in CNTs. This warrants quantitative prediction of the lowest triplet
states in CNTs to be about 0.2-0.3 eV lower than the lowest singlet states, and the triplet state
absorption pronounced at about 0.5-0.6 eV. We further compare properties of calculated singlet and
triplet states in polymers and nanotubes using real-space transition density matrix analysis. This
establishes similarity in electronic properties of these materials and guide design of new
experimental probes of triplet states in CNTs.

\begin{figure*}
  \vspace{-1.6cm}\hspace{-0.6cm} \includegraphics[width=1.03\textwidth]{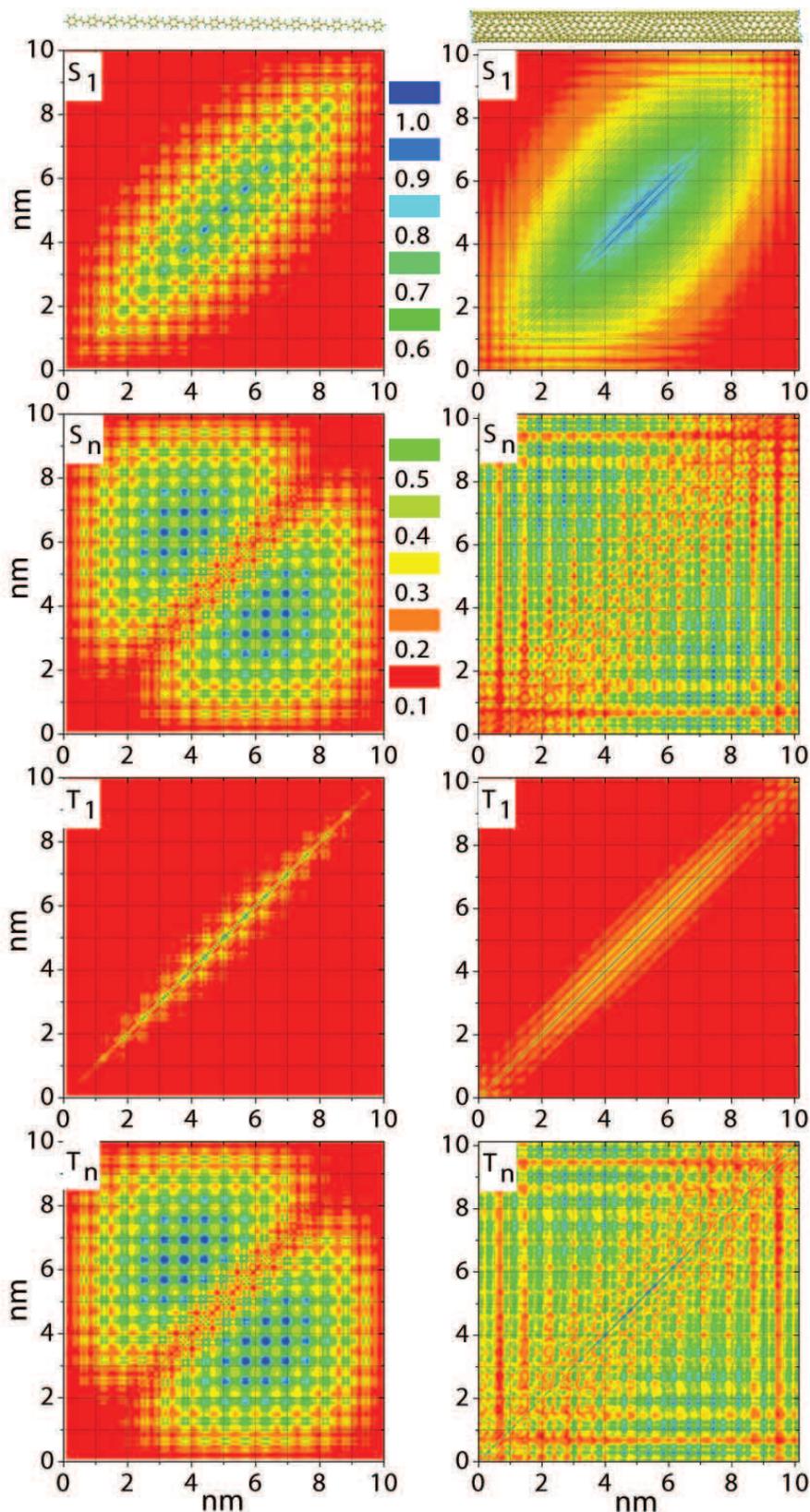} \vspace{-1.5cm}
   \caption{Analysis of transition density matrices corresponding to optically active excited states
   of conjugated polymers (left column) and carbon nanotubes (right column) calculated at PBE1PBE/6-31G level.
   These states are schematically shown in Fig.~\ref{fig:levels}.
   Each plot depicts probabilities of an electron moving from one molecular position (horizontal axis) to
   another (vertical axis) upon electronic excitation. The color code is shown in the middle. }
   \label{fig:modes}
   \end{figure*}

We focus on (7,6) semiconductor tube segment and poly(p-phenylene vinylene) (PPV) oligomer shown in
Fig. \ref{fig:modes} (top). Reference PPV compound is one of the most explored and understood
luminescent polymer \cite{MonkmanAP:Trie}, and (7,6) tubes are common species in experimental
samples. These molecular systems have a finite length of 10 nm and comprise 2 (tube) and 16
(polymer) repeat units. The finite size one-dimensional systems, when their lengths are
significantly larger than the characteristic exciton sizes, are expected to reproduce well the
properties in the infinite-size limit. Calculations of conjugated oligomers provide a standard
example of such an approach \cite{TretiakS:Denmas,TretiakS:Condpc}. Unsaturated chemical bonds at
the CNT ends have been saturated with hydrogen atoms to remove mid-gap states caused by dangling
bonds. The Austin Model 1 (AM1) Hamiltonian \cite{DewarMJS:AM1ngp} has been further used to obtain
optimal geometries of both molecules. This method works very well for geometry optimizations in a
broad variety of $\pi$-conjugated molecular systems. Geometry optimizations conducted with other
methods would lead to small and uniform red (DFT geometries) or blue (Hartree-Fock geometries)
shifts of CNT excitation energies (see supplemental materials). Based on the optimized geometries,
we calculate the corresponding TD-DFT excited state structures up to 25 lowest excited states in
singlet and triplet manifolds using the Gaussian 03 package \cite{g03}. TD-DFT is currently a
mainstream approach for quantitative modeling of optical responses in large molecules
\cite{DreuwA:Sinaim}. We use B3LYP and PBE1PBE hybrid functionals, containing 20\% and 25\% of the
orbital exchange, respectively. These functionals are subjectively considered to be the most
accurate in computational chemistry reproducing well electronic excitations in many materials
\cite{DreuwA:Sinaim}. We emphasize the necessity of using hybrid DFT to account for excitonic
effects \cite{TretiakS.:Excscp}, which are important for molecules in question. Furthermore, the
delicate interplay between energetics of singlet and triplet states and respective exciton binding
energies are extremely sensitive to this exchange component \cite{TretiakS.:Excscp}. Both minimal
valence (STO-3G) and extended (6-31G) basis sets were used for all calculations. Currently such
TD-DFT (e.g. B3LYP/6-31G) calculations became standard in the molecular modeling. However,
computing many excited states in (7,6) tube with more than 1,000 atoms (about 10,000 basis
functions) is extremely numerically demanding and memory intensive task.

Table~\ref{tab:data} summarizes our main computational findings. First of all, we observe several
generic trends expected from the methods used. Minimal basis set (STO-3G) is clearly lacking 'room'
for electronic delocalization. Subsequently, its excitation energies are shifted to the blue
compared to the larger 6-31G basis set results.  We note that the difference is smaller in the
nanotube case. Likely, circumferential dimension of CNT somewhat compensates the reduced basis set
size. Further increase of the basis set (e.g. 6-31G*) will likely have a minor effect on the CNT
excited state energetics leading to red-shifts of excitation energies up to 0.1 eV across the board
(see supplemental materials). Compared to B3LYP, PBE1PBE singlet excitation energies are
blue-shifted. This trend is reversed for the triplet states. This phenomenon is expected as well:
larger fraction of the orbital exchange results in increased shifts of singlet and triplet states
up and down the spectrum, respectively \cite{DreuwA:Sinaim,TretiakS.:Excscp}. Overall, an accuracy
of the calculated excitation energies improves when going from the left to the right in
table~\ref{tab:data}. PBE1PBE/6-31G or B3LYP/6-31G values agree well (within 0.1-0.3 eV) with
available experimental data across the board. Note that the PPV oligomer was calculated in a planar
conformation. Accounting for a torsional disorder typically present in experimental samples would
shift the calculated singlet frequencies to the blue by $\sim$ 0.2 eV improving an agreement with
experiment \cite{TretiakS:Condpc}. Mainly the trend between computed and experimental values is
consistent for all excited state of a given material.

To analyze the excited state properties we use the correspondent transition density matrices,
representing the electronic transition between the ground state and an electronically excited
states \cite{TretiakS:Denmas,TretiakS:Condpc,TretiakS.:Excscp}. The relevant excitonic states of
all molecules are shown in Fig.~\ref{fig:modes}. The matrix diagonal and off-diagonal sizes
characterize the distribution of an excitonic wavefunction over the chain, namely center-of-mass
and distance between particles, respectively. The $S_1$ transition is shown in the first row. Due
to an enhanced excitonic character, the lowest exciton 'collects' most of the oscillator strength
from the band and, subsequently, transition $S_0-S_1$ is strongly optically allowed in both PPV and
(7,6) tube. The contour plots illustrate that the center of mass of photoexcited electron-hole pair
is delocalized over the entire chain (diagonal in the plot) representing typical $\pi-\pi$
excitations. According to the color code, the exciton size (maximal distance between electron and
hole) measured by off-diagonal extent of the non-zero matrix area is about 3 nm in PPV and 4 nm in
nanotube (50\% drop of the wavefunction) and about 5 nm in PPV and 7 nm in nanotube (90\% drop of
the wavefunction). $S_1$ state is the lowest singlet in the PPV oligomers, which exemplifies a
typical case for all photoluminescent polymers. The situation is different in CNTs: our
calculations result in several optically forbidden (or nearly forbidden) exciton states denoted as
D11 in Fig.~\ref{fig:levels}, slightly below (up to 0.1 eV) the allowed $S_1$ state. This relative
ordering of 'bright' and 'dark' states may be responsible for the poor fluorescence efficiency of
the CNTs \cite{ZhaoHB:Eleieo}. These states (not shown) have approximately the same delocalization
properties as $S_1$ state displayed in Fig.~\ref{fig:modes} (top).

$S_n$ (or $mA_g$) state is the next essential state in the singlet excitonic manifolds of both
polymers and nanotubes. This state  shows up as a major peak in both photoinduced (PA1 band) and
two-photon absorptions of nanotubes as evidenced by two recent experimental studies
\cite{ZhaoH.:Phoeqo,WangF:Theorc}. The group theory based on the band model does not predict the
selection rules needed to explain the two photon experiments in CNTs \cite{BarrosEB:Selroa}. Yet,
in the finite molecule calculations presented here, $S_n$ state characteristically appears as a
state with a strong transition dipole moment from $S_1$ state. It is, however, strictly forbidden
in the ground-state absorption. Symmetry notations $1A_g$, $1B_u$, $mA_g$ for $S_0$, $S_1$, and
$S_n$ states, respectively, common in the polymer's photophysics, rationalize analogous appearance
of $S_n$ state in nanotubes as well (a detailed joint experimental/theoretical study appeared
recently \cite{ZhaoH.:Phoeqo}). In the real-space analysis, $S_n$ state is much more delocalized
compared to $S_1$ state with significant spatial separation between an electron and a hole (second
row in Fig.~\ref{fig:modes}), thus representing weakly bound exciton. Subsequently, the energetic
separation between $S_1$ and $S_n$ state has been used as a lower bound estimate of the $S_1$
excitonic binding energy in photoluminescent polymers \cite{OsterbackaR:Excpal} (about 0.6-0.8 eV)
and, recently, in CNTs \cite{ZhaoH.:Phoeqo,WangF:Theorc} (about 0.3-0.4 eV). Above $S_n$ there are
several excited states known in polymer's photophysics as $nB_u$ and $kA_g$, which represent
non-interacting electron-hole pairs at the continuum of the excitonic band. Similar states have
been recently observed in CNTs spectra as well \cite{ZhaoH.:Phoeqo}. All singlet states discussed
above constitute an elegant essential state model applicable for both polymers and nanotubes
\cite{ZhaoH.:Phoeqo}.

Spectroscopic study of triplet states in materials with small spin-orbit coupling is a challenge
for experiment. Yet, the triplet energies has been measured in a broad range of different
$\pi$-conjugated polymers \cite{MonkmanAP:Trie}. This task remains a problem for CNTs. Comparing
the relative energies of the singlet and triplet "gaps" provides an alternative estimate of the
strength of electron-electron correlations, and, particularly, electronic exchange effects.
Typically in one-dimensional materials with strongly bound excitons, there exists a substantial gap
between triplet and singlet energies. For example, the lowest triplet ($T_1$) is about 1 eV lower
than the respective $S_1$ singlet state in conjugated polymers (see Table~\ref{tab:data}). Our
calculations show a similar scenario in CNTs owning to quasi-one-dimensional structure of the
material. In the (7,6) tube there is 0.3-0.4 eV gap between $T_1$ state and optically allowed $S_1$
state. This translates into 0.2-0.3 eV gap between $T_1$ state and optically forbidden lowest
'dark' $D11$ singlet state. We expect that the observed triplet-singlet gap in CNTs will reduce
with increase of the tube diameter, showing behavior similar to the exciton binding energy scaling.
The delocalization pattern of $T_1$ in the (7,6) tube is strikingly similar to that of the PPV
oligomer (see Fig.~\ref{fig:modes}). $T_1$ state in both materials is a tightly bound exciton with
maximal separation between an electron and a hole not exceeding 1 nm. Our results dispute previous
theoretical study based on the empirical model approximation, which assigned triplet states to be
within 20 meV of the singlet bands \cite{PerebeinosV:Radlec}. The latter prediction represents a
typical solid state case and contravenes strong exciton binding energy found in CNTs
\cite{O'ConnellMJ:Bangfi,WangF:Theorc,ZhaoHB:Eleieo,PerebeinosVasili:Scaecn,SpataruCatalinD.:ExcEaO}.

Distinct triplet absorption bands provide a major experimental tool for detection and understanding
of triplet states \cite{MonkmanAP:Trie}. Similar to the singlet manifold with optically allowed
$S_1-S_n$ band, there exists well defined $T_1-T_n$ transition in both polymers and nanotubes.
$T_n$ states have long been explored in the luminescent polymer's context (e.g.
\cite{ZhaoH.:Phoeqo,MonkmanAP:Trie}), where $T_1-T_n$ splitting is about 1.4-2 eV. Our calculations
predict $T_1-T_n$ splitting for (7,6) tube to be about 0.5-0.6 eV (see table~\ref{tab:data}). $T_n$
excitation is a delocalized exciton, which has a structure very similar to that of $S_n$
transition, as evidenced by the transition density plots in Fig.~\ref{fig:modes}. Indeed, with
substantial electron-hole distance, the spin direction becomes of lesser importance compared to the
tightly bound excitons. Consequently, $S_n-T_n$ separation is small compared to the $S_1-T_1$ gap
in conjugated polymers. This $S_n-T_n$ difference becomes even smaller (50 meV) in CNTs as
predicted by our calculations. This gives a powerful clue where to look for $T_n$ state in CNTs
spectroscopies: right below the two-photon allowed $S_n$ excitation.

In conclusion, we have investigated singlet and triplet excited state manifolds of carbon nanotubes
using time-dependent density functional theory. A subset of essential singlet and triplet states,
that dominate photophysical properties of CNTs, is analyzed using calculated transition densities.
These CNT states have very similar properties compared to the analogous states previously observed
in conjugated polymers. Good agreement of our computational results with available experimental
data ensures reliable prediction of triplet energies in CNTs. Our results place CNTs in the same
category of many molecular materials such as acenes, luminescent polymers, and MX chains
\cite{MonkmanAP:Trie,BatistaER:ExcLoc,DexheimerSL:Femvds}, where the energy of the first triplet
state is typically 2/3 that of the first singlet excited state. These evidence significant
electronic correlation effects in CNTs. The lowest CNT triplet excitations $T_1$ are spatially
localized with an excitonic size of about 1 nm. We predict strong optical absorption from $T_1$ to
the higher lying  triplets $T_n$ at about 0.5-0.6 eV. $T_n$ excitations are delocalized states,
which are energetically slightly below the two-photon allowed state $S_n$ in CNTs. These results
provide specific guidelines, which make possible experimental detection of triplet state in CNTs.
The observed subtle interplay between singlet and triplet manifold energetics needs to be accounted
for when designing specific light-driven or electronic nanotechnological applications based on
carbon nanotube materials. We recall that in conjugated polymers triplet states are the dominant
species formed on charge recombination which yields electroluminescence. Substantial deviations
from the spin statistics (i.e., one singlet exciton is formed for every three triplets) favor
singlets and higher luminescence yield, and have been a subject of intense recent debate
\cite{WilsonJS:Spiefp}. Due to spin statics, formation of triplet excitons from the electron and
hole carriers is possible in transport processes, for example, in the CNTs-based chip structures
\cite{PostmaHWC:Carnst,ChenZH:Anilc}. As evidenced by our results, the CNT triplet states have
lower energy and even more tightly bound compared to the singlets. Such excitons can strongly
affect the device performance. Finally, we point out to an interesting possible application of CNT
due to low-lying triplet states: photoprotection against triplet states and singlet oxygen such as
carotenoids functions in photosynthesis \cite{FrankHA:Redfcp}.

We thank A.P. Shreve, A. Piryatinski, G. Lanzani  and S. Kilina  for useful discussions. We
acknowledge support of Center for Nonlinear Studies (CNLS), and LDRD program at LANL. Los Alamos
National Laboratory is operated by Los Alamos National Security, LLC, for the National Nuclear
Security Administration of the U.S. Department of Energy under contract DE-AC52-06NA25396.


\newpage
\begin{table}
\caption{Calculated and experimental excitation energies (eV) of poly-phenylene-vinylene (PPV) and
(7,6) tube. Experimental values have been reported in {\protect
${}^a$\cite{OsterbackaR:Excpal,MonkmanAP:Trie}, ${}^b$\cite{BachiloSM:Stross},
${}^c$\cite{WangF:Theorc,ZhaoH.:Phoeqo}}.} \label{tab:data}
 \vspace{0.2in}
   \begin{tabular}{|c|c|c|c|c|c|c|}\hline\hline
Compound  & Transition& B3LYP   &  PBE1PBE & B3LYP  & PBE1PBE  & Experiment         \\
          &           & STO-3G  &  STO-3G  & 6-31G  & 6-31G    &                    \\ \hline

PPV       & $S_0-S_1$ & 2.80    & 2.99     & 2.24   & 2.37     &  2.48${}^a$        \\
16 units  & $S_0-T_1$ & 1.77    & 1.64     & 1.53   & 1.44     &  1.3${}^a$         \\
          & $S_0-S_n$ & 3.14    & 3.42     & 2.59   & 2.85     &  3.2${}^a$         \\
          & $S_0-T_n$ & 3.05    & 3.32     & 2.50   & 2.75     &  3.0${}^a$         \\\hline
(7,6)     & $S_0-S_1$ & 1.39    & 1.48     & 1.24   & 1.29     &  1.11${}^b$        \\
2 units   & $S_0-T_1$ & 1.05    & 0.95     & 0.97   & 0.94     &   -                \\
          & $S_0-D11$ & 1.29    & 1.37     & 1.15   & 1.20     &  -                 \\
          & $S_0-S_n$ & 1.65    & 1.82     & 1.41   & 1.47     &  $\sim$1.4${}^c$    \\
          & $S_0-T_n$ & 1.63    & 1.78     & 1.39   & 1.45     &  -                 \\ \hline\hline
  \end{tabular}
\end{table}

\end{document}